\documentclass[12pt,a4paper]{article}
\usepackage[english]{babel}
\newtheorem{Theorem}{Theorem}
\newtheorem{Remark}{Remark}
\newtheorem{Lemma}{Lemma}
\newtheorem{Assumption}{A}
\newtheorem{Definition}{Definition}
\newtheorem{Corollary}{Corollary}

\begin{document}
\title{ASYMPTOTICS FOR RELATIVE FREQUENCY WHEN POPULATION IS DRIVEN BY ARBITRARY EVOLUTION}
\author{Silvano Fiorin}
\date{}
\maketitle
\begin{center}
Dipartimento di Scienze Statistiche,via C. Battisti,241-35121 Padova,Italy-fiorin@stat.unipd.it
\end{center}
\newpage
\underline{ABSTRACT} Strongly consistent estimates are shown ,via relative frequency,for the probability of "white balls"
inside a dichotomous urn when such a probability is an arbitrary continuous time dependent function over a bounded time
interval.The asymptotic behaviour of relative frequency is studied in a nonstationary context using a Riemann-Dini type theorem for SLLN of random
variables with arbitrarily different expectations; furthermore the theoretical results concerning the SLLN can be applied
for estimating the mean function of unknown form of a general nonstationary process.
\section{INTRODUCTION}
Several different areas of statistics deal with an urn model including "white" and "black" balls with probability $p$ and 
$1-p$ respectively. In this very classical context a time dependent component is introduced:$p$ is replaced with $p_0(t)$
which denotes a time varying quantity $0\leq p_0(t)\leq 1$ in such a way that at any instant $t\in \lbrack 0,T\rbrack$ only one
observation is taken from the corresponding urn with probability $p_0(t)$ and the random variable $Y(t)$ is obtained such
that $P(Y(t)=1)=p_0(t), P(Y(t)=0)=1-p_0(t),E(Y(t))=p_0(t)$   $\forall t\in\lbrack0,T\rbrack$, defining the nonstationary process
\begin{equation} \label{eq:uno}
\{Y(t):t\in \lbrack0,T\rbrack\}
\end{equation}
with mean function $E(Y(t))=p_0(t)$. The description of the above model is specified introducing some reasonable assumptions:
\newline
\begin{Assumption}\label{eq:quattrocento}
 the continuity is assumed for the usually unknown mean function \newline $p_0:\lbrack0,T\rbrack\mapsto\lbrack0,1\rbrack$
;
\end{Assumption}
\begin{Assumption}\label{eq:quattrocento1}

 for any fixed pair of instants $t_1,t_2 \in \lbrack0,T\rbrack$ the independence is assumed for the random variables
$Y(t_1)$ and $Y(t_2)$.
\end{Assumption}
This assumption is introduced in order to apply the Rajchman theorem(see next section). Namely:only pairwise uncorrelation
is requested for $Y(t_1)$ and $Y(t_2)$ but,it can be easily checked in this case, the uncorrelation implies independence;
furthermore independence is here a very mild condition:in fact we may suppose that the total number of white and black balls
in the urn is big enough that the knowledge of $Y(t_1)=1$ or $Y(t_1)=0$ does not produce a meaningful modification of the
probability distribution for $Y(t_2)$.\newline
The main purpose is estimating the unknown function $p_0$, i.e. the mean function $p_0(t)=E(Y(t))$ of the nonstationary
process (\ref{eq:uno}), which is an arbitrary continuous map form $\lbrack0,T\rbrack$ into $\lbrack0,1\rbrack$.\newline 
i) An approach to estimation for the mean function $m(.)$ of a nonstationary process was given by M.B. Priestley (see \cite{cinque} at page 587 
and \cite{sei} at page 140) when the form of $m$ is known and the case is suggested of a polynomial
function in t. Vice versa :"with no information on the form of $m$ we obviously cannot construct a consistent estimate of 
it". The approach here adopted is quite different from classical methods of time series analysis; the only information 
available for $m$ is the continuity property over $\lbrack 0,T \rbrack$ and no approximatiion of $m$ is introduced by 
continuous functions of a known form. The estimation technique involves the process (\ref{eq:uno}) which is a specified case
of nonstationarity but the theoretical results given in the last section hold true for a general nonstationary process.
The case (\ref{eq:uno}) is only a concrete example of a process having no regularity properties; nevertheless the continuity
for the mean function $m$ is a reasonable and not restrictive assumption which denotes compatibility with a context of an
arbitrary but not brutal evolution for the composition of the urn.\newline
ii) The urn evolution has effects concerning sampling; for instance if the observations number n is big enough a not slight
time interval will be needed in order to receive the n observations which surely are not values taken by the same random
variable. Then, for sake of simplification, we assume that any r.v. $Y(t)$ may be observed at most only one time. The point
of view we adopt is then characterized by a strong nonstationarity and the consistent estimation for the mean $m(t_0)$
at a fixed time $t_0$ may appear as a very hard objective.\newline
iii) The answer to above arguments is the relative frequency 
\begin{equation} \label{eq:due}
\frac{1}{n}\sum_{j=1}^nY(t_j)
\end{equation}
where $\{t_j:j=1,...,n\}$ are the first n observation times of a sequence $\{t_j:j\geq 1\} \subset \lbrack0,T\rbrack$and 
the main purpose is that of getting consistent estimations of $m(t)=p_0(t)$ via almost sure convergence for the sequence
(\ref{eq:due}). The SLLN is then the theoretical tool needed in the below analysis, but the classical approach based on the
zero-mean r.v.'s $(Y(t_j)-p_0(t_j))$, i.e.
\begin{equation} \label{eq:tre}
\frac{1}{n}\sum_{j=1}^n(Y(t_j)-p_0(t_j)) \to 0 \textrm{ a.s.}
\end{equation}
is not enough:in fact we need convergence for (\ref{eq:due}) with the not zero mean r.v.'s $Y(t_j)$. This argument,
investigated by Fiorin \cite{quattro} is now improved with the help of new results given in section (5).\\
iv)The convergence of (\ref{eq:due}) is studied via the sequence $\{E(Y(t_j))=p_0(t_j):j\geq1 \}$ and permutations 
(i.e. bijections) $\pi:N \to N$:in fact, if a permutation $\pi$ is introduced, the possible almost sure limit of
\begin{equation} \label{eq:quattro}
\frac{1}{n}\sum_{j=1}^nY(t_{\pi(j)})
\end{equation}
is depending on $\pi$. If $\{P_{\pi n}^0\}$ is a sequence of probability measures,where each $P_{\pi n}^0$ assigns mass
$\frac{1}{n}$ to each point $\{p_0(t_{\pi(j)}):j=1,...,n\}$, then the "weak" or "vague" convergence for the sequence 
$\{P_{\pi n}^0\}$ to a probability measure $P^0$ implies almost sure convergence of (\ref{eq:quattro}) to the limit
$\int_{0}^{1}I(v) dP^0(v)$ where $I(v)$ is the idntity map over $\lbrack0,1\rbrack$ and $P^0$ depends on the sequence
$\{Y(t_j):j\geq1\}$ and on permutation $\pi$. All the below analysis is based on the possibility of finding a permutation
$\pi$ in such a way that the convergence of (\ref{eq:quattro}) is driven to a limit $\int_{0}^{1} I(v) dP^0(v)$ where $P^0$
is a previously chosen probability measure over $\lbrack0,1\rbrack$;under a theoretical point of view this is a result for
SLLN (\ref{eq:quattro}) which is the analogous of the well known Riemann-Dini theorem for real simply convergent (but not
absolutely convergent) series. Under the operative point of view the strongly consistent estimates, i.e. the a.s. limits
$\int_{0}^{1} I(v) dP^0(v)$, are the result of an experimental design based on choosing:\\
I) the sequence of observation times $\{t_j:j\geq1\} \subset \lbrack0,T\rbrack$;\\
II) the permutation $\{t_{\pi(j)}:j\geq1\}$.

\section{CONVERGENCE ELEMENTS}
If the observation times $\{t_j:j\geq1\}$ are given jointly with the observable r.v.'s $\{Y(t_j):j\geq 1\}$, an intuitive
approach for studying the almost sure convergence for (\ref{eq:due}) is suggested by the classical Rajchman theorem
\begin{Theorem}\label{eq:cento1}
If the $Y(t_j)$'s are pairwise uncorrelated and their second moments have a common bound then 
\begin{displaymath}
\frac{1}{n}\sum_{j=1}^n(Y(t_j)-p_0(t_j))
\end{displaymath}
is convergent to 0 almost surely.
\end{Theorem}
Because of assumption A2) and the inequality $|Y(t_j)|\leq1$ the $Y(t_j)$'s satisfy theorem (\ref{eq:cento1}) and then
\begin{equation}\label{eq:cinque}
\frac{1}{n}\sum_{j=1}^n(Y(t_j)-p_0(t_j))\to0 \textrm{ } a.s..
\end{equation}
Now an intuitive and simple condition which implies (together with (\ref{eq:cinque})) the almost sure convergence for 
(\ref{eq:due}) is the possible limit for the deterministic sequence 
\begin{equation}\label{eq:sei}
\frac{1}{n}\sum_{j=1}^np_0(t_j)=\frac{1}{n}\sum_{j=1}^nE(Y(t_j)).
\end{equation}
In fact,if such a limit exists,i.e.
\begin{displaymath}
L=\lim_{n\to\infty}\frac{1}{n}\sum_{j=1}^np_0(t_j),
\end{displaymath}
we have
\begin{displaymath}
\frac{1}{n}\sum_{j=1}^nY(t_j)\to L \textrm{ } a.s..
\end{displaymath}
\begin{Definition}\label{eq:2cento1}
Let us define as a "pseudoempirical measure" (P.E.M. hereafter) any probability measure giving the weight $\frac{1}{n}$ to
each of the assigned points $\{x_j:j=1,....,n\}$, where the "pseudo" means that the $x_j$'s are arbitrarily fixed
deterministic values and not a sequence of i.i.d. observations.
\end{Definition}
The notion of "Vague Convergence" (V.C. hereafter) is introduced mainly for application to sequences of P.E.M.'s;
such a concept,which implies existence of limit L for the sequence (\ref{eq:sei}),is the main technical tool for studying
the asymptotic behaviour of relative frequency (\ref{eq:due}).Only the really necessary elements for below analysis are
here given;for an exhaustive exposition see Chung \cite{tre}.
\begin{Definition}\label{eq:2cento2}
A sequence $\{\mu_n:n\geq1\}$ of probability measures (P.M. hereafter) defined over the Borel $\sigma$-field $B^1$
of $R^1$ is said to converge vaguely to the P.M. $\mu$ iff there exists a dense subset $D$ of $R^1$ such that
\begin{displaymath}
\mu_n(a,b]\to \mu(a,b],\forall a\in D,b\in D,a< b.
\end{displaymath}  
\end{Definition}
\begin{Theorem}\label{eq:cento2}
(see Theorem 4.3.1,page 85 Chung \cite{tre})The sequence of P.M.'s $\mu_n$ is vaguely convergent to the P.M. $\mu$ if and only if
\begin{displaymath}
\lim_{n\to\infty}\mu_n(a,b]=\mu(a,b]
\end{displaymath}
for every continuity interval $(a,b]$ of $\mu$,i.e. for every interval whose endpoints satisfy $\mu(a)=\mu(b)=0$.
\end{Theorem}
By theorem (\ref{eq:cento2})the equivalence is stated between vague and weak convergence for P.M.'s $\mu_n$ to $\mu$.
A further classical result needed in the below proofs is the following characterization of V.C.:
\begin{Theorem}\label{eq:cento3}
(see Theorem 4.4.2.,page 93 Chung \cite{tre}) $ \mu_n$ is vaguely convergent to $\mu$ if and only if the convergence is stated
\begin{displaymath}
\lim_{n\to\infty} \int_{R} f d\mu_n =\int_{R} f d\mu
\end{displaymath}
for each bounded,continuous and real  f.
\end{Theorem}
Even if vague and weak convergence of P.M.'s are equivalent,in the main proofs the V.C. is preferable because the 
convergence has to be proved $\mu_n(a,b]\to \mu(a,b]$ for countably many a,b in a dense subset of $R$.\newline
The above theorem (\ref{eq:cento3}) can be directly applied for convergence of sequence (\ref{eq:sei}) via the equality
\begin{equation}\label{eq:sette}
\frac{1}{n}\sum_{j=1}^np_0(t_j)=\int_{0}^{1} p  dP_n^0
\end{equation}
where $P_n^0$ is the P.E.M. giving weight $\frac{1}{n}$ to each point $\{p_0(t_j):j=1,...,n\}$.Thus a condition which 
implies the convergence of (\ref{eq:sei}) is the vague convergence for the sequence of P.E.M.'s $P_n^0$ to a P.M. $P^0$.
In fact if $P_n^0$ is V.C. to $P^0$,having $p_0(t_j)\in[0,1] \forall j$,and taking the function
\begin{displaymath}
f(p)=p \textrm{ } \forall p \in \lbrack0,1\rbrack
\end{displaymath}
\begin{displaymath}
f(p)=1 \textrm{ } \forall p \in \lbrack1,+\infty)
\end{displaymath}
\begin{displaymath}
f(p)=0 \textrm{ } \forall p \in (-\infty,0\rbrack,
\end{displaymath}
by theorem (\ref{eq:cento3}),the convergence is stated
\begin{displaymath}
\int_{0}^{1} p dP_n^0 \to \int_{0}^{1} p dP^0
\end{displaymath}
i.e.
\begin{equation}\label{eq:otto}
\frac{1}{n}\sum_{j=1}^np_0(t_j)\to\int_{0}^{1}p dP^0
\end{equation}
which jointly with theorem (\ref{eq:cento1}) implies
\begin{equation}\label{eq:nove}
\frac{1}{n}\sum_{j=1}^nY(t_j)\to \int_{0}^{1} p dP^0 a.s..
\end{equation}
\begin{Remark}\label{eq:3cento1}
For the almost sure convergence (\ref{eq:nove}) an alternative proof is given by theorem (\ref{eq:cento6}) below:working with the sequence
$\frac{1}{n}\sum_{j=1}^nY(t_j)$ its direct approximation to the integral $\int_{0}^{1} p dP^0$ is proved.
\end{Remark}
The central argument concerning convergence (\ref{eq:nove})is the assumption of vague convergence for $P_n^0$ to $P^0$.
Several questions may arise:for instance it is evident that such a condition is not so easy to reach.In fact the 
restrictivity of this assumption will be evident via Definition (\ref{eq:2cento2}):for an assigned sequence of expextations
$\{E(Y(t_j))=p_0(t_j):j\geq1\}$ and a fixed interval $(a,b\rbrack\subset\lbrack0,1\rbrack$ the convergence 
$P_n^0(a,b\rbrack \to P^0(a,b\rbrack$ holds true where
\begin{displaymath}
P_n^0(a,b\rbrack = \frac{n(a,b\rbrack}{n}
\end{displaymath}
and $n(a,b\rbrack$ is the total number of points $\{p_0(t_j):j=1,...,n\}$ belonging to $(a,b\rbrack$:this means that 
inside the first n elements of the sequence $\{p_0(t_j):j\geq1\}$ the proportion of ponts falling into $(a,b\rbrack$
is "so regular" to approach a limit $P^0(a,b\rbrack$,when $n \to \infty$.And this for an arbitrary deterministic sequence
$\{p_0(t_j):j\geq1\}\subset\lbrack0,1\rbrack$.\newline
Our purpose ,in the sequel, will consist of a strategy to obtain a vaguely convergent sequence of P.E.M.'s $P_n^0$;
recalling I) and II) at the end of introduction,we may choose an experimental design which consists of two steps;
we may decide when to observe the continuous time process$\{Y(t):t\in \lbrack0,T\rbrack\}$ and then we choose the 
observation times consisting of a sequence $\{t_j:j\geq1\}$ $\subset$ $\lbrack0,T\rbrack$.Not only:we may decide also,
for each n fixed,the n observable r.v.'s to choose inside $\{Y(t_j):j\geq1\}$,i.e. we do not consider necessarily the 
first n r.v.'s $\{Y(t_j):j=1,...,n\}$ but we select $\{Y(t_{\pi(j)}:j=1,...,n\}$ with the respective expectations
$\{E(Y(t_{\pi(j)}):j=1,...,n\}$ where $\{\pi(j):j=1,...,n\}$ are the first n values taken by a permutation (a bijection)
$\pi$:N$\to$N,in such a way that,if $P_{\pi n}$ denotes the P.E.M. giving mass $\frac{1}{n}$ to each point 
$\{t_{\pi(j)}:j=1,...,n\}$,the sequence $P_{\pi n}$ is vaguely or weakly convergent to some P.M. $P_{\pi}$.Then,using the
relevant property that the induced measures $p_0(P_{\pi n})$'s and $p_0(P_{\pi})$ keep the weak convergence, we reach the
V.C. $p_0(P_{\pi n}) \to p_0(P_{\pi})$, where $p_0(P_{\pi n})$ assigns mass $\frac{1}{n}$ to each point
$\{p_0(t_{\pi(j)}):j=1,...,n\}$.But,for a complete description of the above strategy,we need to introduce the relevant 
tool of permutations.
\section{PERMUTATIONS}
Given the family of r.v.'s $\{Y(t_j):j\geq1\}$ with expectations $\{E(Y(t_j))=p_0(t_j):j\geq1\}$,for any assigned bijection
$\pi$:N$\to$N the respective process may be defined
\begin{displaymath}
\{Y(t_{\pi(j)}):j\geq1\}
\end{displaymath}
with expectations $\{E(Y(t_{\pi(j)})=p_0(t_{\pi(j)}):\j\geq1\}$ and the P.E.M.'s $P_{\pi n}^0$ which gives mass $\frac{1}{n}$
to each point $\{p_0(t_{\pi(j)}):j=1,...,n\}$.A direct comparison shows that $P_n^0$ and $P_{\pi n}^0$,in the general case,
define different probability measures over the Borel $\sigma$-field $B\lbrack0,1\rbrack$.Consequently the possible vague
limits $P^0$ and $P_{\pi}^0$,if they exist,are different P.M.'s and,applying Theorem (\ref{eq:cento3}) and Theorem (\ref{eq:cento6}),we have:
\begin{displaymath}
\frac{1}{n}\sum_{j=1}^np_0(t_j)\to \int_{0}^{1} p dP^0 \textrm{ and } \frac{1}{n}\sum_{j=1}^nY(t_j)\to \int_{0}^{1} p dP^0 a.s.
\end{displaymath}
and using permutation $\pi$
\begin{displaymath}
\frac{1}{n}\sum_{j=1}^np_0(t_{\pi(j)})\to \int_{0}^{1} p dP_{\pi}^0 \textrm{ and }
\frac{1}{n}\sum_{j=1}^nY(t_{\pi(j)})\to \int_{0}^{1} pdP_{\pi}^0 a.s..
\end{displaymath}
Permutations are an important argument in below analysis with several implications concerning estimation;then this topic 
needs further attention:the vague convergence for a sequence of P.M.'s $P_{\pi n}^0$ was introduced above only as an
hypothesis.Now,in order to obtain an estimation procedure,the following three steps have to be examined:\newline
1)the vague convergence for an assigned sequence of P.E.M.'s $P_{\pi n}^0$ has really to be proved.\newline
2)Given the sequence of points $\{t_j:j\geq1\} \subset \lbrack0,T\rbrack$,the class $\mathcal{M}$ has to be found of P.M.'s
P over $B\lbrack0,T\rbrack$ for which a permutation $\{t_{\pi(j)}:j\geq1\}$ can be computed such that the P.E.M.'s 
$P_{\pi n}$ (which assigns weight $\frac{1}{n}$ to each point $\{t_{\pi(j)}:j=1,...,n\}$) are vaguely convergent to P
and then the induced measures $p_0(P_{\pi n})$ over $B\lbrack0,1\rbrack$ are vaguely convergent to $p_0(P)$(because of
continuity of $p_0$),where
\begin{displaymath}
p_0(P_{\pi n})(B)=P_{\pi n}(p_0^{-1}(B))\textrm{ and } p_0(P)(B)=P(p_0^{-1}(B)) \forall B\in B\lbrack0,1\rbrack.
\end{displaymath}
3)The possibility of choosing a measure $P\in\mathcal{M}$, and then of computing a permutation $\{t_{\pi(j)}:j\geq1\}$
such that ,applying theorem (\ref{eq:cento6}),
\begin{displaymath}
\frac{1}{n}\sum_{j=1}^nY(t_{\pi(j)})\to \int_{0}^{1} p d p_0(P)\textrm{ }a.s.,
\end{displaymath}
is a good chance for consistent estimation:through the choice of the vague limit measure P and of $\pi$ the convergence 
for the SLLN may be driven to different limit values.\newline
A rigorous characterization of class $\mathcal{M}$ is given by definition (\ref{eq:2cento6}) which needs more technical details given later;
nevertheless it may be useful to anticipate the content of assumption under which $\mathcal{M}$  contains infinitely many
measures:if the set of points $\{t_j:j\geq1\}\subset\lbrack0,T\rbrack$ has at least two different limit values,i.e.
if there are at least two values $L_1 \neq L_2$ such that there exist two subsequences 
\begin{displaymath}
\lim_{k\to\infty}t_{j_1(k)}=L_1 \textrm{ and } \lim_{k\to\infty}t_{j_2(k)}=L_2,
\end{displaymath}
then $\mathcal{M}$ contains infinitely many probability measures.\newline
Furthermore ,for an assigned measure $P\in\mathcal{M}$, the procedure of finding a permutation $\{t_{\pi(j)}:j\geq1\}$such
that the respective P.E.M.'s $P_{\pi n}$ are vaguely convergent to the assigned P is available in the proof of theorem (\ref{eq:cento7}).
Our aim consists now in applying the above results for estimation.
\section{ESTIMATING $p_0$}
As examples of estimation problems two different procedures are shown below where suitable choices of the sequences of
obsevatioin times $\{t_j:j\geq1\}$ and of permutations $\{t_{\pi(j)}:j\geq1\}$ imply almost sure convergence for SLLN
$\frac{1}{n}\sum_{j=1}^nY(t_{\pi(j)})$ to different estimations.
\subsection{PROBLEM 1}
Let us suppose to choose a sequence of observation times $\{t_j:j\geq1\}$ which is dense into $\lbrack0,T\rbrack$;then
by Corollary (\ref{eq:one}) the class $\mathcal{M }$ contain the uniform probability measure $P_U$ over 
$B\lbrack0,T\rbrack$ which is characterized by the respective density function $f_U(t)=\frac{1}{T} \forall t\in\lbrack0,T\rbrack$
and ,applying the proof of theorem (\ref{eq:cento7}) a permutation $\{t_{\pi(j)}:j\geq1\}$ is computed such that the P.E.M.'s $P_ {\pi n}$
are vaguely convergent to $P_U$.Now,for a fixed interval $(a,b\rbrack\subset\lbrack0,T\rbrack$ and for any assigned natural
n,the following set is introduced:
\begin{displaymath}
A(\pi,n,(a,b\rbrack)=\{t_{\pi(j)}\in(a,b\rbrack:j=1,...,n\}
\end{displaymath}
whose meaning is evident:among the points $\{t_{\pi(j)}:j=1,...,n\}$ only the $t_{\pi(j)}$'s falling inside $(a,b\rbrack$
are collected.If $n(a,b\rbrack$ is the total number of points $t_{\pi(j)}$'s belonging to $A(\pi,n,(a,b\rbrack)$ and the
relative frequency is introduced
\begin{equation}\label{eq:dieci}
\frac{1}{n(a,b\rbrack}\sum_{t_{\pi(j)}\in A(\pi,n,(a,b\rbrack)}Y(t_{\pi(j)})
\end{equation}
the a.s. convergence for (\ref{eq:dieci}) ,when $n\to\infty$ and then necessarily $n(a,b\rbrack \to \infty$, is stated by
below theorem
\begin{Theorem}\label{eq:cento4}
The sequence of r.v.'s (\ref{eq:dieci}),when $n\to \infty$ is a strongly consistent estimate of $p_0(\underline{t})$ for some points 
$\underline{t} \in \lbrack a,b\rbrack$.
\end{Theorem}
\underline{Proof of Theorem}
By Corollary (\ref{eq:one})to main Theorem (\ref{eq:cento7}) a permutation $\{t_{\pi(j)}:j\geq1\}$ can be found such that the P.E.M.'s $P_{\pi n}$ are
vaguely convergent to the uniform measure $P_U$ (with density function $f_U(t)=\frac{1}{T} \forall t \in \lbrack0,T\rbrack$),
where $P_{\pi n}$ assigns mass $\frac{1}{n}$ to each point $\{t_{\pi(j)}:j=1,...,n\}$;thus, for each fixed interval
$(a,b\rbrack\subset\lbrack0,T\rbrack$, we have

\begin{displaymath}
P_{\pi n}(a,b\rbrack=\frac{n(a,b\rbrack}{n}\textrm{ and }\lim_{n\to\infty}\frac{n(a,b\rbrack}{n}=P_U(a,b\rbrack=\frac{b-a}{T}
\end{displaymath}

and this because each $(a,b\rbrack$ is a $P_U$-continuity set.Now for a fixed $(a,b\rbrack$ let us denote by
$P_{(\pi n(a,b\rbrack)}$ the probability measure giving mass $\frac{1}{n(a,b\rbrack}$ to each point $t_{\pi(j)}\in A(\pi,n,(a,b\rbrack)$
in such a way that
\begin{displaymath}
P_{(\pi n(a,b\rbrack)}(c,d\rbrack=\frac{n(c,d\rbrack}{n(a,b\rbrack}\textrm{ } \forall (c,d\rbrack\subset(a,b\rbrack,
\end{displaymath}
where $n(c,d\rbrack$ is defined analogously to $n(a,b\rbrack$.Let us observe that,because of the equality
\begin{displaymath}
\frac{n(c,d\rbrack}{n(a,b\rbrack}=\frac{n(c,d\rbrack/n}{n(a,b\rbrack/n}=P_{\pi n}(c,d\rbrack\frac{1}{P_{\pi n}(a,b\rbrack},
\end{displaymath}
and the vague convergence $P_{\pi n} \to P_U$,we have
\begin{displaymath}
\lim_{n\to \infty}\frac{n(c,d\rbrack}{n(a,b\rbrack}=\lim_{n\to\infty}P_{\pi n}(c,d\rbrack \frac{1}{\lim_{n\to\infty}
P_{\pi n}(a,b\rbrack}=\frac{d-c}{T}\frac{T}{b-a}=\frac{d-c}{b-a},
\end{displaymath}
i.e.
\begin{displaymath}
\lim_{n\to\infty}P_ {(\pi n(a,b\rbrack)}(c,d\rbrack=\frac{d-c}{b-a}
\end{displaymath}
and the sequence of P.E.M.'s $P_{(\pi n(a,b\rbrack)}$ is vaguely convergent to uniform measure $P_{U(a,b\rbrack}$
having density function $f_{U(a,b\rbrack}(t)=\frac{1}{b-a} \forall t \in (a,b\rbrack$.Denoting with
$p_0(P_{(\pi n(a,b\rbrack)})$ and $p_0(P_{U(a,b\rbrack})$ the induced measures by $p_0$,i.e.
\begin{displaymath}
p_0(P_{(\pi n(a,b\rbrack))}(B)=P_{(\pi n(a,b\rbrack)}(p_0^{-1}(B))\textrm{and}p_0(P_{U(a,b\rbrack})(B)=P_{U(a,b\rbrack}
(p_0^{-1}(B)),
\end{displaymath}
$\forall B\in B\lbrack0,1\rbrack$;because of continuity of $p_0$,the vague convergence of $P_{(\pi n(a,b\rbrack)}$ to 
$P_{U(a,b\rbrack}$ implies the vague convergence of $p_0(P_{(\pi n(a,b\rbrack)})$ to $p_0(P_{U(a,b\rbrack})$ and then,
by Theorem (\ref{eq:cento6}),the convergences hold true
\begin{displaymath}
\lim_{n \to \infty}\frac{1}{n(a,b\rbrack}\sum_{t_{\pi(j)} \in A(\pi,n,(a,b\rbrack)}p_0(t_{\pi(j)})= 
\int_{0}^{1} p dp_0(P_{U(a,b\rbrack})
\end{displaymath}
and
\begin{displaymath}
\frac{1}{n(a,b\rbrack}\sum_{t_{(\pi(j)}\in A(\pi , n, (a,b\rbrack)}Y(t_{\pi(j)}) \to \int_{0}^{1} p dp_0(P_{U(a,b\rbrack})a.s.;
\end{displaymath}
finally,by standard analysis arguments,
\begin{displaymath}
\int_{0}^{1} p dp_0(P_{U(a,b\rbrack})=\int_{a}^{b} p_0(t) dP_{U(a,b\rbrack}=\int_{a}^{b} p_0(t)\frac{1}{b-a} dt=
\end{displaymath}
\begin{displaymath}
\frac{1}{b-a}p_0(\underline{t})(b-a)=p_0(\underline{t})
\end{displaymath}
where $\underline{t}$ is a point whose existence is stated by the mean value Theorem for integral of the continuous
$p_0$ function and proof is now complete.
\subsection{PROBLEM 2}
Our interest is now concerning a strongly consistent estimate of $p_0(t)$ where $t\in\lbrack0,T\rbrack$ is assigned.The
elementary solution given by $\frac{1}{n}\sum_{i=1}^nY_i(t)$ and based on the observations$Y_1(t),...,Y_n(t)$ of the r.v.
$Y(t)$ has no meaning in our context; in fact we may suppose that,when n is big enough,taking n observations at the same
instant t is not possible and then we necessarily need n observation instants $t_1,t_2,...,t_n$ with the respective 
r.v.'s $Y(t_1),Y(t_2),...,Y(t_n)$ and their expectations $p_0(t_1),p_0(t_2),...,p_0(t_n)$, and this because our urn model 
has a time dependent composition.\newline
Our aim consists in proving the following result:
\begin{Theorem}\label{eq:cento5}
If $\{t_j:j\geq1\}$ is any convergent sequence to t,then $\frac{1}{n}\sum_{j=1}^nY(t_j)$ is a strongly consistent estimate
of $p_0(t)$.
\end{Theorem}
Proof of Theorem.A first elementary proof is given proving that the convergence $p_0(t_j)\to p_0(t)$ implies convergence
$\frac{1}{n}\sum_{j=1}^np_0(t_j) \to p_0(t)$. In fact ,because of convergence $p_0(t_j) \to p_0(t)$, for fixed
$\frac{\epsilon}{2}$ there exists k such that $|p_0(t_j)-p_0(t)|<\frac{\epsilon}{2} \forall j>k$,and then
\begin{displaymath}
\frac{1}{n}\sum_{j=1}^np_0(t_j)=\frac{1}{n}\sum_{j=1}^kp_0(t_j)+\frac{n-k}{n}\frac{\sum_{j=k+1}^np_0(t_j)}{n-k}
\end{displaymath}
and 
\begin{displaymath}
|\frac{\sum_{j=k+1}^np_0(t_j)}{n-k}-p_0(t)|=|\frac{\sum_{j=k+1}^np_0(t_j)}{n-k}-\frac{(n-k)p_0(t)}{n-k}|\leq
\end{displaymath}
\begin{displaymath}
\frac{1}{n-k}\sum_{j=k+1}^n|p_0(t_j)-p_0(t)|\leq \frac{1}{n-k}(n-k)\frac{\epsilon}{2}=\frac{\epsilon}{2} \forall n>k.
\end{displaymath}
Finally the limits $\frac{1}{n}\sum_{j=1}^kp_0(t_j) \to 0$ and $\frac{n-k}{n} \to 1$, when $n\to \infty$,allows us  to state
the existence of $n_0$ such that
\begin{displaymath}
|\frac{1}{n}\sum_{j=1}^np_0(t_j)-p_0(t)|<\epsilon \textrm{ } \forall n>n_0;
\end{displaymath}
proving that $\lim_{n \to \infty}\frac{1}{n}\sum_{j=1}^np_0(t_j)=p_0(t)$,which jointly with the almost sure convergence
$\frac{1}{n}\sum_{j=1}^n(Y(t_j)-p_0(t_j)) \to 0$ (apply Rajchman Theorem) completes the proof.\newline
The same result may be proved also via vague convergence of P.E.M.'s $P_n^0$ which assigns 
 weight $\frac{1}{n}$ to each
point $\{p_0(t_j):j=1,...,n\} \forall n $ fixed.If $(a,b\rbrack$ is an interval having t as an internal point,then there
exists k such that $p_0(t_j)\in (a,b\rbrack \forall j>k$ and $P_n^0(a,b\rbrack \to 1$,while if t is internal to the
complement of $(a,b\rbrack$ we have that $P_n^0(a,b\rbrack \to 0$,proving that $P_n^0$ is vaguely convergent to 
$P=\delta_t$ which assigns weight 1 to point t.Applying Theorem (\ref{eq:cento6}) the result is proved.\newline
Applying again the above technique a consistent estimation is found for the difference
\begin{displaymath}
\lbrack p_0(t)-p_0(t^-)\rbrack \textrm{where} p_0(t^-)=\lim_{s\to t^-}p_0(s)
\end{displaymath}
if the function $p_0$ is right continuous with left limits.In fact if $\{t_j:j\geq1\}$ and $\{s_j:j\geq1\}$ are two sequences
satisfying
\begin{displaymath}
\lim_{j\to \infty} t_j=t^+ \textrm{and} \lim_{j\to \infty} s_j=t^-
\end{displaymath}
we have
\begin{displaymath}
\lim_{j \to \infty} p_0(t_j)=p_0(t) \textrm{and} \lim_{j \to \infty} p_0(s_j)=p_0(t^-),
\end{displaymath}
thus applying the above Theorem (\ref{eq:cento5}) we obtain
\begin{displaymath}
\frac{1}{n}\sum_{j=1}^nY(t_j) \to p_0(t) \textrm{and} \frac{1}{n}\sum_{j=1}^nY(s_j) \to p_0(t^-) \textrm{ } a.s.
\end{displaymath}
and then
\begin{displaymath}
\frac{1}{n}\sum_{j=1}^nY(t_j)-\frac{1}{n}\sum_{j=1}^nY(s_j) \to \lbrack p_0(t)-p_0(t^-)\rbrack \textrm{ } a.s..
\end{displaymath}
\section{A RIEMANN-DINI TYPE THEOREM FOR SLLN}
The well known Riemann-Dini theorem for real numbers series is extended to strong laws of large numbers for real random
variables.Namely:if $\sum_{j=1}^{\infty}x_j$ is a simply but not an absolutely convergent series of real numbers and 
$\alpha\in \mathcal{R} \cup \{\infty,-\infty\}$ is an assigned value,then there exists a permutation (i.e. a bijection
$\pi$:N$\to$N) such that $\sum_{j=1}^{\infty}x_{\pi(j)}=\alpha$.Analogously,given a sequence of real random variables
$\{Y_j:j\geq1\}$ having arbitrarily different and finite expectations $\{E(Y_j):j\geq1\}$,it is shown,under suitable 
assumptions,that for any fixed real number $\beta$ belonging to a wide class $B\subset \mathcal{R}$,there exists a 
permutation $\pi$:N $\to$ N such that the sequence $\frac{1}{n}\sum_{j=1}^nY_{\pi(j)}$ is almost surely convergent to
$\beta$ when $n \to \infty$.The main technical tool is the study of convergence for the sequences of measures $P_n$ which 
assigns probability mass $\frac{1}{n}$ to each value $\{E(Y_j):j=1,...,n \}$ and of the deep interplay between the possible
limits of sequences $\{P_n:n\geq1\}$ and the permutations of values $\{E(Y_{\pi(j)}):j\geq1\}$ where $\pi$:N$ \to $N is an
assigned bijection. 
\subsection{PRELIMINARY ELEMENTS}
As an introductory argument a simple but meaningful example may help in showing the goal of our analysis.\newline
\underline{EXAMPLE 1}\newline
Let us suppose that there exists a partition for the sequence of real r.v.'s $\{Y_j:j\geq1\}$ into two subsequences
denoted by $\{Y_{l_{k}}:k\geq1\}$ and $\{Y_{n_{k}}:k\geq1\}$ satisfying
\begin{equation}\label{eq:undici}
\{Y_j:j\geq1\}=\{Y_{l_{k}}:k\geq1\} \cup \{Y_{n_{k}}:k\geq1\}
\end{equation}
where $E(Y_{l_{k}})=L_1,E(Y_{n_{k}})=L_2,\forall k\geq1$.\newline
For each fixed natural n let $C_n(L_1)$ and $C_n(L_2)$ denote respectively the total number of r.v.'s $Y_j$ with
$1\leq j \leq n$ which satisfy $E(Y_j)=L_1$ or $E(Y_j)=L_2$ in such a way that $n=C_n(L_1)+C_n(L_2)$ and then
\begin{displaymath}
\sum_{j=1}^nY_j=\sum_{k=1}^{C_n(L_1)}Y_{l_{k}}+\sum_{K=1}^{C_n(L_2)}Y_{n_{k}}.
\end{displaymath}
Consequently we obtain
\begin{equation}\label{eq:dodici}
\frac{1}{n}\sum_{j=1}^nY_j=\frac{C_n(L_1)}{n}\frac{\sum_{k=1}^{c_n(L_1)}Y_{l_{k}}}{C_n(L_1)}+\frac{C_n(L_2)}{n}
\frac{\sum_{K=1}^{C_n(L_2)}Y_{n_{k}}}{C_n(L_2)}
\end{equation}
where
\begin{equation}\label{eq:tredici}
0\leq \frac{C_n(L_1)}{n} \leq 1, 0\leq \frac{C_n(L_2)}{n} \leq 1 \textrm{ and } \frac{C_n(L_1)}{n}+\frac{C_n(L_2)}{n}=1.
\end{equation}
Because of (\ref{eq:dodici}) the convergence for $\frac{1}{n}\sum_{j=1}^nY_j$ can be shown if the following two steps 
procedure holds true:\newline
a)applying the standard SLLN the convergences are stated
\begin{displaymath}
\frac{\sum_{k=1}^{C_n(L_1)}Y_{l_{k}}}{C_n(L_1)} \to L_1 \textrm{ and } \frac{\sum_{k=1}^{C_n(L_2)}Y_{n_{k}}}{C_n(L_2)} \to L_2
\end{displaymath}
almost surely when $n \to \infty$;\newline
b)if
\begin{equation}\label{eq:quattordici}
\lim_{n \to \infty}\frac{C_n(L_1)}{n}=p_1,\lim_{n \to \infty}\frac{C_n(L_2)}{n}=p_2,
\end{equation}
because of (\ref{eq:tredici}) $p_1+p_2=1$ and then the pair $(p_1,p_2)$ defines a probability distribution over the real
values $L_1,L_2$.Then,under a) and b) above,we have
\begin{equation}\label{eq:quindici}
\frac{1}{n}\sum_{j=1}^ny_j \to p_1L_1+p_2L_2 \textrm{almost surely.}
\end{equation}
Now this simple case allows us to detect the main elements of our analysis:\newline
i)a class of limit values $p_1L_1+p_2L_2$ can be introduced for fixed $L_1$ and $L_2$ when the pair $(p_1,p_2)$ is 
arbitrarily chosen under conditions $0\leq p_i \leq 1$ for $i=1,2$ and $p_1+p_2=1$ in such a way that for fixed $L_1$ and
$L_2$ the set
\begin{equation}\label{eq:sedici}
B(L_1,L_2)=\{p_1L_1+p_2L_2:0\leq p_i \leq 1  (i=1,2),  p_1+p_2=1\}
\end{equation}
defines all possible values which can be the almost sure limit for a sequence
\begin{displaymath}
\frac{1}{n}\sum_{j=1}^nY_{\pi(j)} \textrm{where  } \pi: N \to N \textrm{ is a permutation of } Y_j's.
\end{displaymath}
ii)the existence is evident of a strict connection between any fixed value $p_1L_1+p_2L_2 \in B(L_1,L_2)$ and a permutation
$\pi$ such that $\frac{1}{n}\sum_{j=1}^nY_{\pi(j)} \to p_1L_1+p_2L_2$.\newline
iii)the almost sure limit $p_1L_1+p_2L_2$ can be written as an integral
\begin{equation}\label{eq:diciassette}
\int_{R} I(v) d(p_1\delta_{L_1}+p_2\delta_{L_2})
\end{equation}
where $I(.)$ is the identity map and $p_1\delta_{L_1}+p_2\delta_{L_2}$ is the probability measure giving mass $p_1$ to
$L_1$ and $p_2$ to $L_2$ respectively. This measure is defined through the strict interplay of two components:\newline
c1)the values $L_1$ and $L_2$ which are assigned by the expextations $E(Y_j)$'s;\newline
c2)the probability distribution denoted with $p_1$ and $p_2$ which is the result of limits (\ref{eq:quattordici}) and
choosing a permutation of $Y_j$'s.Such a probability measure plays a central role in our approach:for any fixed pair
$(p_1,p_2)$ with $0\leq p_i \leq 1$ and $p_1+p_2=1$ there exists some permutations $\pi$ such that 
\begin{displaymath}
\frac{1}{n}\sum_{j=1}^nY_{\pi(j)} \to \int_{R} I(v) d(p_1\delta_{L_1}+p_2\delta_{L_2})\textrm{ } a.s.;
\end{displaymath}
thus the limit for the SLLN is assigned by measure $p_1\delta_{L_1}+p_2\delta_{L_2}$. $\diamondsuit$  \newline
Our aim consists in extending the above example 1 to more general situations:for instance if the set $\{E(Y_j):j\geq1\}$
contains arbitrarily different values,including the case when $E(Y_j) \neq E(Y_k) \textrm{ } \forall j\neq k$.The main result deals
with a sequence of r.v.'s $\{Y_j:j\geq1\}$ under the following assumptions:
\begin{Assumption}\label{eq:quattrocento2}
the $Y_j$'s are uniformly bounded i.e. there exists $M>0$ such that $|Y_j|\leq M \textrm{ } \forall j \geq1$;
\end{Assumption}
\begin{Assumption}\label{eq:quattrocento3}
the $Y_j$'s are pairwise uncorrelated;
\end{Assumption}
\begin{Assumption}\label{eq:quattrocento4}
the $Y_j$'s have probability distributions and finite expectations which are arbitrarily different;
\end{Assumption}
\begin{Assumption}\label{eq:quattrocento5}
the sequence of expectations $\{E(Y_j):j\geq1\} \subset \lbrack-M,M\rbrack$ has at least two different limit points,
i.e. there exist at least two different values $x_1,x_2 \in \lbrack -M,M\rbrack$ which are the limits of some 
subsequences of $\{E(Y_j):j\geq1\}$.
\end{Assumption}
It will be shown below the existence of a wide class $\mathcal{M}$ of probability measures P over the Borel $\sigma$-field
$B(-M,M\rbrack$ such that for any assigned $P \in \mathcal{M}$  there exist some permutations $\pi$:N $\to$ N satisfying
\begin{equation}\label{eq:diciotto}
\frac{1}{n}\sum_{j=1}^nY_{\pi(j)} \to \int_{-M}^{M} I(v) d P(v) \textrm{ } a.s..
\end{equation} 
The representation of limits given in (\ref{eq:diciotto}) by integrals of type $\int_{-M}^{M} I(v) d P(v)$ gives big
evidence to measure P;not only: the convergence stated by (\ref{eq:diciotto}) and the approach here adopted are mainly
based on measures defined over the interval$(-M,M\rbrack$.Namely:P is a probability measure which is the limit in some
sense of the sequence of the P.E.M. $P_{\pi n}$'s which assigns weight $\frac{1}{n}$ to each point $\{E(Y_{\pi(j)}):j=1,...,n\}$;
moreover, if the permutation $\pi$ is adopted, the set of mean values $\{E(Y_{\pi(j)}):j=1,...,n\}$, the P.E.M.'s 
$P_{\pi n}$, and the possible limit P depend on $\pi$. The detailed and rigorous definition of the class $\mathcal{M}$
needs several technical elements which will be an argument of the below subsections.\newline
A further intuitive argument may help in understanding the meaning of our aim; if the r.v.'s $\{Y_j:j\geq1\}$ satisfy
above assumptions and have arbitrarily different expectations $\{E(Y_j):j\geq1\}$ a SLLN can be easily given taking the
differences $\{(Y_j-E(Y_j)):j\geq1\}$ and then applying a well known result:see,for instance, theorem 5.1.2 at page 108
of Chung book \cite{tre} .In fact the $Y_j$'s are uncorrelated and with uniformly bounded second moments ,then
\begin{equation}\label{eq:diciannove}
\frac{1}{n}\sum_{j=1}^n(Y_j-E(Y_j))=\frac{1}{n}\sum_{j=1}^nY_j-\frac{1}{n}\sum_{j=1}^nE(Y_j) \to 0 \textrm{ } a.s..
\end{equation}
Of course this is not a solution to our problem :the (\ref{eq:diciannove}) in fact states the convergence to 0 for the 
differences and a convergence result for $\frac{1}{n}\sum_{j=1}^nE(Y_j)$ is not so easy to obtain.A law of large numbers
cannot be applied to the deterministic sequence $\{E(Y_j):j\geq1\}$ and also the convergence for the series
$\sum_{j=1}^{\infty}\frac{E(Y_j)}{j}$, in order to apply the Kronecker lemma, is not an easy one if $\{E(Y_j):j\geq1\}$ is
a general sequence in the interval $(-M,M \rbrack$. On the other hand the convergence for $\frac{1}{n}\sum_{j=1}^nE(Y_j)$
to some value L jointly with the SLLN (\ref{eq:diciannove}) implies that $\frac{1}{n}\sum_{j=1}^nY_j \to L$ a.s. solving 
our problem. That of finding hypotheses under which the sequence $\frac{1}{n}\sum_{j=1}^nE(Y_j)$ is a convergent one is
then a relevant tool in this context. Let us write $\frac{1}{n}\sum_{j=1}^nE(Y_j)$ as an integral,i.e.
\begin{equation}\label{eq:venti}
\frac{1}{n}\sum_{j=1}^nE(Y_j)=\int_{-M}^{M}I(v) dP_n(v)=\int_{R}I_M(v) dP_n(v)
\end{equation}
where $P_n$ is the P.E.M. giving mass $\frac{1}{n}$ to each point $\{E(Y_j):j=1,...,n\}$, $I(v)$ is the identity map and
\begin{displaymath}
I_M(v)= I(v) \textrm{ if } v \in (-M,M)
\end{displaymath}
\begin{displaymath}
I_M(v)= M \textrm{ if } v \in \lbrack M,\infty)
\end{displaymath}
\begin{displaymath}
I_M(v)= -M \textrm{ if } v \in (-\infty,-M \rbrack.
\end{displaymath}
Because of continuity and boundedness of $I_M$ over $R^1$ a favourable context for convergence of the integrals sequence
\begin{displaymath}
\{\int_{R^1} I_M(v) dP_n(v) :n\geq1\}
\end{displaymath}
is given by VAGUE CONVERGENCE for the sequence $\{P_n\}$ of probability measures. Applying Theorem 4.4.2 at page 93 of 
Chung book \cite{tre}  we have that if $P_n,P$ are probability measures, then $\{P_n\}$ is vaguely convergent to $P$ if and only if
\begin{displaymath}
\int_{R^1} f(v) dP_n(v) \to \int_{R^1} f(v) dP(v)
\end{displaymath}
for each continuous and bounded f.\newline
Thus the vague convergence of $P_n$'s to $P$ implies convergence for integrals
\begin{displaymath}
\int_{R^1} I_M(v) dP_n(v) \to \int_{R^1} I_M(v) dP(v),
\end{displaymath}
thus
\begin{displaymath}
\lim_{n \to \infty}\frac{1}{n}\sum_{j=1}^nE(Y_j)= \int_{-M}^{M} I(v) dP(v) \textrm{ and}
\end{displaymath}
\begin{displaymath}
\lim_{n \to \infty} \frac{1}{n}\sum_{j=1}^nY_j= \int_{-M}^{M} I(v) dP(v) \textrm{ a.s..}
\end{displaymath}
The vague convergence of P.E.M. $P_n$'s is the general setting adopted for our analysis: the centrality of its role,now
evident for convergence of $\frac{1}{n}\sum_{j=1}^nE(Y_j)$, will be shown below also for directly proving the convergence
of $\frac{1}{n}\sum_{j=1}^nY_j$.
\subsection{THE TECHNICAL BACKGROUND}
For a fixed natural m let us denote by $\mathcal{H}_m=\{H_r:r=1,...,m\}$ a partition of the interval $(-M,M\rbrack$ into m
 subintervals where
\begin{equation}\label{eq:ventuno}
H_1=(-M,t_1\rbrack, H_2=(t_1,t_2\rbrack,...,H_m=(t_{m-1},M\rbrack;
\end{equation}
the sequence of r.v.'s $\{Y_j:j\geq1\}$ is supposed to satisfy Assumption (\ref{eq:quattrocento2}),...,
Assumption (\ref{eq:quattrocento5}) and a permutation $\pi$, which is assigned for $Y_j$'s, is omitted in the notations in
order to semplify formulas.A partition for $\{Y_j:j\geq1\}$ into a family of m subsequences is introduced on the base of 
the m sets $\{H_r:r=1,...,m\}$:for each fixed $H_r$ we collect the $Y_j$'s having the respective $E(Y_j) \in H_r$, i.e.
the subsequence is introduced
\begin{equation}\label{eq:ventidue}
\{Y_{jr_{k}}:k=1,2,...,Q(H_r)\}
\end{equation}
where:\newline
i)Q is the counting measure which assigns to each $B \in B(-M,M\rbrack$ the respective value $Q(B)$ i.e. the total
number of values $E(Y_j) \in B$.Thus the set of values taken by Q includes any natural n and also $+\infty$.\newline
ii)The index $jr$ is a strictly increasing map $jr:N \to N$ and any value $jr_k=jr(k)$ means that $Y_j$ with 
$j=jr_k$ is the k-th element inside $\{Y_j:j\geq1\}$ such that $E(Y_j) \in H_r$.Thus each of the m subsequences 
$\{Y_{jr_{k}}:k=1,...,Q(r)\}$ with $r=1,...,m$ is characterized through the respective index, i.e. the strictly increasing
map $jr:N \to N$,satisfies the following properties:\newline
I)the m sets of values $\{jr_k=jr(k):k=1,2,...,Q(r)\}$ for $r=1,...,m$ are pairwise disjoint;\newline 
II)their union is equal to N.\newline
Then the m subsequences $\{Y_{jr_{k}}:k=1,...,Q(r)\}$ for $r=1,...,m$ are a partition of $\{Y_j:j\geq1\}$.Now ,for each
fixed natural n and given $\{Y_j:j=1,...,n\}$ and $\{E(Y_j):j=1,...,n\}$,let us define the quantities
$\{c_n(r):r=1,...,m\}$as 
\begin{equation}\label{eq:ventitre}
C_n(r)=\sum_{j=1}^n I_{H_{r}}(E(Y_j))
\end{equation}
where $I_{H_{r}}(E(Y_j))=1$ if $E(Y_j) \in H_r$ and $I_{H_{r}}(E(Y_j))=0$ if $E(Y_j) \notin H_r$; $C_n(r)$ is then the total
number of values in the set $\{E(Y_j):j=1,...,n\}$ falling inside the interval $H_r$.The following quantity is a
generalization of (\ref{eq:dodici}) concerning EXAMPLE 1
\begin{equation}\label{eq:ventiquattro}
\frac{1}{n}\sum_{j=1}^nY_j=\sum_{r=1}^m\frac{C_n(r)}{n}\frac{\sum_{k=1}^{C_n(r)}Y_{jr_{k}}}{C_n(r)}.
\end{equation}
A technical tool for below proofs consisits in studying the limit for the second member of (\ref{eq:ventiquattro}) when
$n \to \infty$.A two step procedure is pointed out dealing, for a fixed r, with the two sequences $\frac{C_n(r)}{n}$ and
$\frac{\sum_{k=1}^{C_n(r)}Y_{jr_{k}}}{C_n(r)}$.Of course the interesting case is when $H_r$ contains infinitely many 
$E(Y_j)$'s and then $\frac{C_n(r)}{n}$ may be convergent to a non zero limit.\newline
\underline{STEP 1} The convergence is assumed 
\begin{displaymath}
P(H_r)= \lim_{n \to \infty} \frac{C_n(r)}{n} \textrm{ } \forall r=1,...,m,
\end{displaymath}
where $P$ is an assigned probability measure over the Borel $\sigma$-field $B(-M,M \rbrack$.\newline
\underline{STEP 2} If $H_r$ includes infinitely many values $E(Y_j)$'s,then the SLLN can be applied to the sequence 
\begin{equation}\label{eq:venticinque}
\frac{1}{C_n(r)}\sum_{k=1}^{C_n(r)}Y_{jr_{k}}=\frac{1}{C_n(r)}\sum_{k=1}^{C_n(r)}(Y_{jr_{k}}-E(Y_{jr_{k}}))+
\frac{1}{C_n(r)}\sum_{k=1}^{C_n(r)}E(Y_{jr_{k}}).
\end{equation}
Because of Assumptions (\ref{eq:quattrocento2}) and (\ref{eq:quattrocento3}) the SLLN (see Theorem 5.1.2 of Chung book \cite{tre}  is
applied to the first term in second member of (\ref{eq:venticinque})
\begin{displaymath}
\frac{1}{n}\sum_{k=1}^{C_n(r)}(Y_{jr_{k}}-E(Y_{jr_{k}})) \to 0 \textrm{  a.s..}
\end{displaymath}
The inclusion $E(Y_{jr_{k}}) \in H_r=(t_{r-1},t_r\rbrack$ means $t_{r-1}<E(Y_{jr_{k}})\leq t_r$ and then the below inequality
\begin{equation}\label{eq:ventisei}
t_{r-1}<\frac{\sum_{k=1}^{C_n(r)}E(Y_{jr_{k}})}{C_n(r)} \leq t_r
\end{equation}
states that the oscillations of the sequence $\frac{1}{C_n(r)}\sum_{k=1}^{C_n(r)}E(Y_{jr_{k}})$ can be made arbitrarily small
if the length of $H_r$ is small and the above steps imply that
\begin{equation}\label{eq:ventisette}
|\frac{C_n(r)}{n}\frac{\sum_{k=1}^{C_n(r)}Y_{jr_{k}}}{C_n(r)}-P(H_r)\frac{\sum_{k=1}^{C_n(r)}E(Y_{jr_{k}})}{C_n(r)}| \to 0
\textrm{ a.s.}
\end{equation}
and
\begin{equation} \label{eq:ventotto}
|\sum_{r=1}^m\frac{C_n(r)}{n}\frac{\sum_{k=1}^{C_n(r)}Y_{jr_{k}}}{C_n(r)}-\sum_{r=1}^m P(H_r)\frac{\sum_{k=1}^{C_n(r)}
E(Y_{jr_{k}})}{C_n(r)}| \to 0 \textrm{ a.s..}
\end{equation}
We are now ready for the below statement:
\begin{Lemma} \label{eq:cinquecento}
If the sequence of r.v.'s $\{Y_j:j\geq1\}$ satisfies Assumptions (\ref{eq:quattrocento2}),(\ref{eq:quattrocento3}) and if,
for $\epsilon$ fixed,there exists a partition of $(-M,M\rbrack$ into subsets $\{H_r:r=1,...,m\}$ such that:\newline
i)the length of each $H_r$ is not grater than $\epsilon$;\newline
ii)$\lim_{n \to \infty}\frac{C_n(r)}{n}=P(H_r)\textrm{ } \forall r=1,...,m$ where $P$ is an assigned probability measure
over $B(-M,M\rbrack$, then there exists a set $A$ with probability one such that for each $\omega \in A$
the existence is proved of a natural value $n_0(\epsilon,\omega)$ satisfying
\begin{displaymath}
|\frac{1}{n}\sum_{j=1}^nY_j(\omega)-\int_{-M}^{M}I(v) dP(v)|<2\epsilon, \textrm{ }\forall n>n_0(\epsilon,\omega).
\end{displaymath}
\end{Lemma}
\underline{PROOF OF LEMMA (\ref{eq:cinquecento})} . The sequence of r.v.'s $\{Y_{jr_{k}}:k\geq1\}$ satisfies
Assumptions (\ref{eq:quattrocento2}) and (\ref{eq:quattrocento3}) and then,applying Theorem 5.1.2 of Chung book \cite{tre}  ,the existence
is proved for a set $A_r \subset \Omega$ with $\mu(A_r)=1$,where $\mu$ is the probability measure defined over $\Omega$, such that
\begin{displaymath}
\frac{1}{C_n(r)}\sum_{K=1}^{C_n(r)}(Y_{jr_{k}}-E(Y_{jr_{k}})) \to 0
\end{displaymath}
over the set $A_r$.Of course the above arguments are concerning a set $H_r$ including infinitely many values $E(Y_j)$'s
in such a way that $C_n(r) \to \infty$ when $n \to \infty$;now using (\ref{eq:venticinque}), the convergence 
(\ref{eq:ventisette}) can be directly proved.Through iterations of above procedure for each $r=1,...,m$ the existence is
given of sets $\{A_r:r=1,...,m\}$  with $\mu(A_r)=1$ $\forall r=1,...,m$ and then through the intersection
$A= \cap_{r=1}^m A_r$ we have that $\mu(A)=1$ and (\ref{eq:ventotto}) holds true.If the value 
\begin{displaymath}
\sum_{r=1}^mP(H_r)\frac{\sum_{k=1}^{C_n(r)}E(Y_{jr_{k}})}{C_n(r)}
\end{displaymath}
is thought as the integral of a simple function taking a constant value over each interval $H_r$, than its distance from
$\int_{-M}^{M} I(v) dP(v)$ can be estimated using standard arguments:
\begin{displaymath}
|\sum_{r=1}^mP(H_r)\frac{\sum_{k=1}^{C_n(r)}E(Y_{jr_{k}})}{C_n(r)}-\int_{-M}^{M} I(v) dP(v)|\leq
\end{displaymath}
\begin{displaymath}
\leq \sum_{r=1}^m\ \int_{H_r} |\frac{\sum_{k=1}^{C_n(r)}E(Y_{jr_{k}})}{C_n(r)}-I(v)| dP(v)\leq
\end{displaymath}
\begin{displaymath}
\leq \sum_{r=1}^m \epsilon P(H_r)= \epsilon \sum_{r=1}^mP(H_r)= \epsilon
\end{displaymath}
and this recalling that $\frac{\sum_{k=1}^{C_n(r)}E(Y_{jr_{k}})}{C_n(r)} \in H_r$ $\forall r=1,...,m$ and if the length
of each $H_r$ is at most $\epsilon$. The result follows from (\ref{eq:ventotto}) and the last inequalities.
\subsection{THE MEASURES $P_n$,$P$,$Q$}
Two types of measures introduced above  have a central role:\newline
1)the counting measure $Q$ (see (\ref{eq:ventidue})), whose values $Q(B)$ assigns the total number of $E(Y_j)$'s falling 
into $B$, $\forall B \in B(-M,M\rbrack$; $Q$ is based on the position of $E(Y_j)$'s inside $(-M,M\rbrack$ and may take any
natural value and $+ \infty$ too.\newline
2) keeping on account of $Q$, and a fixed permutation $\pi$ for $Y_j$'s and $E(Y_j)$'s,the quantities $\frac{C_n(r)}{n}$
were introduced (see (\ref{eq:ventitre}) and (\ref{eq:ventiquattro})) for each $H_r$ with $r=1,...,m$; $C_n(r)$ is the 
total number of values $E(Y_j)$'s belonging to $H_r$.If a different permutation $\pi$' is chosen for $Y_j$'s,different
values $C'_n(r)$ will be generated.Thus if the limit exists $\lim_{n \to \infty}\frac{C'_n(r)}{n}=P'(H_r)$
$\forall r=1,...,m$,where $P'$ is a probability measure over $B(-M,M\rbrack$, the $P'$ depends on measure $Q$ and 
permutation $\pi$'.A more general way to define the quantities $\frac{C_n(r)}{n}$ is that of introducing the probability
measure $P_n$ which assigns the mass $\frac{1}{n}$ to each value $\{E(Y_j):j=1,...,n\}$ for n fixed and then
\begin{equation} \label{eq:ventinove}
P_n=\sum_{j=1}^n\frac{1}{n} \textrm{ } \delta_{E(Y_j)}
\end{equation}
defines a probability measure over $B(-M,M\rbrack$ which is referred as "pseudoempiric measure" (P.E.M.) where the
"pseudo" means that $\{E(Y_j):j\geq1\}$ is a deterministic and not an i.i.d. sequence of observations.The above limits,
if they exist, may be rewritten as
\begin{equation} \label{eq:trenta}
\lim_{n \to \infty} P_n(H_r)=P(H_r) \textrm{ } \forall r=1,...,m
\end{equation}
and the close interplay between permutation $\pi$ and measure $P$ is one of the interesting aspects which characterize
the context with arbitrarily different expectations $E(Y_j)$'s,where the P.E.M. $P_n$'s and the possible limit measure
$P$ are strictly dependent on $\pi$.If $E(Y_j)=v_0$ $\forall j\geq1$,i.e. if we consider the classical case, then we have
$P_n=P=\delta_{v_{0}}$, and this for any assigned permutation $\pi$ showing that the classical case is invariant with
respect to permutations.
\subsection{THE CONVERGENCE OF $P_n$'s TO P}
This subsection deals mainly with the type of convergence to adopt for the sequence of P.E.M. $P_n$'s to $P$.Each $P_n$ and
$P$ are defined over the Borel $\sigma$-field $B(-M,M\rbrack$ and then it may appear as a natural request to ask that the
convergence $\lim_{n \to \infty} P_n(B)=P(B)$ holds true for each $B \in B(-M,M\rbrack$.The following example shows that 
convergence $P_n(B) \to P(B)$ $\forall B \in B(-M,M\rbrack$ is a too restrictive request for our purposes.\newline
\underline{EXAMPLE 2}\textrm{ } Let us suppose that $\{E(Y_j):j\geq1\}$ is a strictly decreasing sequence inside $(-M,M\rbrack$ such
that $L=\lim_{j \to \infty} E(Y_j)$ and then a sequence of intervals
\begin{displaymath}
A_j=(a_j,b_j\rbrack \subset \lbrack-M,M\rbrack \textrm{ } \forall j\geq1
\end{displaymath}
can be constructed in such a way that\newline
i)$A_j$ contains only one $E(Y_j)$ as an internal point;\newline
ii)$A_j \cap A_l = \emptyset$ $\forall j \neq l $.
A permutation $\pi$ is assigned and the corresponding sequence $\{Y_{\pi(j)}:j\geq1\}$ is considered; for each n fixed let
$P_n$ be the P.E.M. which assigns probability mass $\frac{1}{n}$ to each point $\{E(Y_{\pi(j)}):j=1,...,n\}$ and then
$P_n(A_{\pi(j)})=\frac{1}{n}$ if $j=1,...,n$ and $P_n(A_{\pi(j)})=0$ if $j>n$.Because of the equalities
\begin{displaymath}
\sum_{j=1}^{\infty}P_n(A(_{\pi(j)})=1 \textrm{ } \forall n \textrm{ fixed and } \lim_{n \to \infty}P_n(A_{\pi(j)})=0
\textrm{ }\forall j \textrm{ fixed,}
\end{displaymath}
the Steinhaus Lemma (see Ash book \cite{uno}  at page 44 ) ensures the existence of a subsequence $\{A_{\pi(j_{k}}):k\geq1\}$ such that
$\{P_n(\cup_{k=1}^{\infty} A_{\pi(j_{k})}):n\geq1\}$ is not a convergent sequence,proving that the convergence $P_n(b)
 \to P(B)$ does not hold true over all sets of $B(-M,M\rbrack$, and this for any assigned permutation $\pi$.$\diamondsuit$ \newline
Now the above example 2 suggests to adopt a type of convergence $P_n \to P$ which is based on a suitable subclass of
$B(-M,M\rbrack$:then the VAGUE CONVERGENCE of $P_n$ to $P$ is considered as a driving element for main results given below.
The general definition (see Chung book \cite{tre}  at page 85) is given when $P_n$,$P$ are subprobability measures;nevertheless,in this
context,we are dealing only with probability measures and then we prefer to consider this case. Moreover,as
$P_n(-M,M\rbrack=P(-M,M\rbrack=1$ we may suppose,without loss of generality,to handle probability measures $P$ satisfying
$P(-M)=P(M)=0$ and $Q(-M)=Q(M)=0$,where $Q$ is the counting measure.The above elements suggest us to use a condition for 
vague convergence of probability measures which is equivalent to the general one over $R^1$ but using only the interval
$(-M,M\rbrack$.Some preliminary notions are needed to introduce the definition of vague convergence given below.
In (\ref{eq:ventuno}) we denoted as $\mathcal{H}_m=\{H_r:r=1,...,m\}$ a partition of $(-M,M\rbrack$ into m subintervals
$H_1=(-M,t_1\rbrack, H_2=(t_1,t_2\rbrack, ...,H_m=(t_{m-1},M\rbrack$.\newline
Here a sequence of partitions for $(-M,M\rbrack$ is introduced as it follows; $\mathcal{H}_1=(-M,M\rbrack$ contains 
$(-M,M\rbrack$ as its unique element.Then choosing arbitrarily a point $t_3$ satisfying $-M<t_3<M$ the partition
$\mathcal{H}_2$ is obtained consisting of two intervals $\mathcal{H}_2=\{(-M,t_3\rbrack,(t_3,M\rbrack\}$ and choosing
$t_4$ such that $-M<t_4<t_3$ the partition $\mathcal{H}_3$ consists of three intervals $\mathcal{H}_3=\{(-M,t_4\rbrack,
(t_4,t_3\rbrack,(t_3,M\rbrack\}$.If $t_5$ is chosen with $t_3<t_5<M$ we have $\mathcal{H}_4=\{-M,t_4\rbrack,(t_4,t_3\rbrack,
(t_3,t_5\rbrack,(t_5,M\rbrack\}$ and so on..... generating a sequence of partitions $\{\mathcal{H}_m:m\geq1\}$.
\begin{Definition}\label{eq:2cento3}
A sequence of partitions $\{\mathcal{H}_m:m\geq1\}$ generated by above procedure is defined to be a "progressive sequence
of partitions" (P.S.P. hereafter) if $\lim_{m \to \infty} l_m=0$ where $l_m$ is the maximum length of the m intervals 
included into $\mathcal{H}_m$.
\end{Definition}
\begin{Definition} \label{eq:2cento4}
An interval $(a,b\rbrack$ is defined to be a continuity interval for the probability measure $P$ defined over the Borel
$\sigma$-field $B(R^1)$ if $P(a)=P(b)=0$.
\end{Definition}
\begin{Definition} \label{eq:2cento5}
If $P_n$,$P$ are probability measures satisfying $P(-M)=P(M)=P_n(-M)=P_n(M)=0$ and $P(-M,M\rbrack=P_n(-M,M\rbrack=1$,the
sequence $\{P_n\}$ is defined to be vaguely convergent to $P$ if there exists a P.S.P. $\{\mathcal{H}_m:m\geq1\}$ such that
each interval $H \in \cup_{m} \mathcal{H}_m$ is a continuity interval for $P$ and $\lim_{n \to \infty} P_n(H)=P(H)$.
\end{Definition}
\subsection{THE MAIN RESULTS}
Let us suppose that the sequence of P.E.M. $P_n$'s,satisfying Definition (\ref{eq:2cento5}),is vaguely convergent to $P$.
Thus the convergence holds true $\lim_{n \to \infty}P_n(H)=P(H)$ for each $H$ inside  a P.S.P. $\cup_m\mathcal{H}_m$ and
consequently if $Q(H)=k \in N$ ($k<+\infty$),then 
\begin{equation} \label{eq:trentuno}
Q(H)=k \Rightarrow P(H)=0
\end{equation}
or equivalently
\begin{equation} \label{eq:trentadue}
P(H)>0 \Rightarrow Q(H)=+\infty
\end{equation}
where $Q(H)$ is the counting measure which assigns the total number of values $E(Y_j) \in H$.Condition (\ref{eq:trentuno})
seems to be very close to absolute continuity of $P$ with respect to $Q$;nevertheless the absolute continuity is defined 
over the Borel $\sigma$-field $B(-M,M\rbrack$ while (\ref{eq:trentuno}) involves only intervals inside $\cup_m\mathcal{H}_m$
where $\{\mathcal{H}_m:m\geq1\}$ is a P.S.P..In our context conditions (\ref{eq:trentuno}) or (\ref{eq:trentadue}) are more
general than absolute continuity $P<<Q$;the evidence is reached via some simple examples,and this could be the case when
$\{E(Y_j):j\geq1\}$ is a convergent sequence to L and $E(Y_j) \neq L \textrm{ } \forall j\geq1$.If L is an interior point 
of $(a,b\rbrack$ then $\lim_{n \to \infty}P_n(a,b\rbrack=1$ and $\lim_{n \to \infty}P_n(a,b\rbrack=0$ if L is interior to
the complement of $(a,b\rbrack$.Denoting as $P=\delta_L$ the probability measure giving mass 1 to L,a P.S.P. 
$\{\mathcal{H}_m:m\geq1\}$ for $(-M,M\rbrack$ is easy to obtain such that each $H \in \cup_m\mathcal{H}_m$ is a $P$
-continuity set and $\lim_{n \to \infty}P_n(H)=P(H)$.Now $P(H)>0$ means $P(H)=1$ and this implies $Q(H)=+\infty$ showing
that (\ref{eq:trentadue}) holds true.Nevertheless,being $E(Y_j)\neq L \textrm{ } \forall j$ we have $Q(L)=0$ and
$P(L)=1$ showing that $P$ is not absolutely continuous with respect to $Q$ (over the Borel $\sigma$-field) and this even if
(\ref{eq:trentuno}) and (\ref{eq:trentadue}) hold true over a P.S.P. $\{\mathcal{H}_m:m\geq1\}$ of $P$-continuity sets.
Condition (\ref{eq:trentuno}) or (\ref{eq:trentadue}) has a central role in main results described by the following two
statements.\newline
Our interest is concerning an assigned sequence of r.v.'s $\{Y_j:j\geq1\}$ with finite expectations $\{E(Y_j):j\geq1\}$
satifying Assumptions (\ref{eq:quattrocento2})-(\ref{eq:quattrocento5});the P.E.M. $P_n$ gives mass $\frac{1}{n}$ to each 
of n values $\{E(Y_j):j=1,...,n\}$ and $Q$ is the counting measure defined above.
\begin{Theorem}\label{eq:cento6}
If the sequence of P.E.M. $P_n$'s is vaguely convergent to a probability measure $P$,then the convergence is satisfied
\begin{displaymath}
\frac{1}{n}\sum_{j=1}^nY_j  \to \int_{-M}^{M} I(v) dP(v) \textrm{ a.s..}
\end{displaymath}
\end{Theorem}
Of course $P$ satisfies (\ref{eq:trentuno}) and (\ref{eq:trentadue}) with respect to $Q$ because of vague convergence of
$P_n$'s to $P$. Such a relationship shows its importance also in the main statement which is ,in some sense, the converse
of above Theorem (\ref{eq:cento6}):given a probability measure $P$ over $B(-M,M\rbrack$ does exist a condition which ensures
the existence of a permutation $\pi:N \to N$ such that $\frac{1}{n}\sum_{j=1}^n Y_{\pi(j)} \to \int_{-M}^{M} I(v) dP(v) 
\textrm{ a.s.?}$ The answer is (\ref{eq:trentadue}):using such a condition the class $\mathcal{M}$ is introduced.
\begin{Definition} \label{eq:2cento6}
Given the sequence of r.v.'s $\{Y_j:j\geq1\}$ with finite expectations $\{E(Y_j):j\geq1\}$ satisfying Assumptions
(\ref{eq:quattrocento2})-(\ref{eq:quattrocento5}), let $\mathcal{M}$ denotes the class of probability measures $P$ over
$B(-M,M\rbrack$ having a P.S.P. $\{\mathcal{H}_m:m\geq1\}$ of P-continuity sets such that $P(H)>0 \Rightarrow Q(H)= +\infty
\textrm{ } \forall H \in \cup_m\mathcal{H}_m$.
\end{Definition}
\begin{Theorem}\label{eq:cento7}
For each assigned probability measure $P \in \mathcal{M}$ a permutation $\pi:N \to N$ can be computed such that the sequence
of P.E.M. $P_{\pi n}$'s (which for each n fixed assigns mass $\frac{1}{n}$ to each value $\{E(Y_{\pi(j)}:j=1,...,n\}$) is
vaguely convergent to $P$ and then (by Theorem (\ref{eq:cento6}))
\begin{displaymath}
\frac{1}{n}\sum_{j=1}^nY_{\pi(j)} \to \int_{-M}^{M} I(v) dP(v) \textrm{ a.s..}
\end{displaymath}
\end{Theorem}
\subsection{PROOF OF MAIN RESULTS}
\underline{PROOF OF THEOREM (\ref{eq:cento6})} Applying definitions (\ref{eq:2cento3}),(\ref{eq:2cento4}),(\ref{eq:2cento5})
there exists a P.S.P. $\{\mathcal{H}_m:m\geq1\}$ of P-continuity sets such that\newline
i)$ \lim_{n \to \infty}P_n(H)=P(H)\textrm{ } \forall H \in \cup_m \mathcal{H}$;\newline
ii)$\lim_{m \to \infty}\epsilon_m=0$ where $\epsilon_m$ is the maximum length of the set of intervals
$\{H_r:r=1,...,m\}=\mathcal{H}_m$.\newline
Then ,applying Lemma (\ref{eq:cinquecento}) to each fixed partition $\mathcal{H}_m$, the existence is shown for a set $A_m$
such that \newline
a)$\mu(A_m)=1$ where $\mu$ is the  probability measure defined over $\Omega$.\newline
b)for each $\omega  \in A_m$, there exists an integer $n_0(\epsilon_m,\omega)$ such that
\begin{displaymath}
|\frac{1}{n}\sum_{j=1}^nY_j(\omega)-\int_{-M}^{M} I(v) dP(v)|<2\epsilon_m \textrm{ } \forall n>n_0(\epsilon_m,\omega).
\end{displaymath}
Thus the $\mu(\cap_{m=1}^{\infty} A_m)=1$ and each $\omega \in \cap_m A_m$,satisfying statement b) above 
for each $m\geq1$, shows the convergence $\frac{1}{n}\sum_{j=1}^nY_j(\omega) \to \int_{-M}^{M} I(v) dP(v)$ a.s..\newline
\underline{PROOF OF THEOREM (\ref{eq:cento7})}The starting point is a probability measure $P$ over $B(-M,M\rbrack$ which
admits a P.S.P. $\{\mathcal{H}_m:m\geq1\}$ of P-continuity sets $H$'s such that $P(H)>0 \Rightarrow Q(H)=\infty$
$\forall H \in \cup_m \mathcal{H}_m$.The below proof consists of several steps.\newline
\underline{1)THE STRUCTURE OF PARTITIONS}\newline
Recalling the construction for the P.S.P. $\{\mathcal{H}_m:m\geq1\}$,the partition $\mathcal{H}_m$ is a class of m right
closed and left open intervals $H_{rm} \subset (-M,M\rbrack$ indexed by rm i.e. $\mathcal{H}_m=\{H_{rm}:rm=1,2,...,m\}$
and inside $\mathcal{H}_m$ we separate the sets $H_{rm}$ having positive and null P-measure:
\begin{displaymath}
\mathcal{H}_m^+=\{H_{rm}:P(H_{rm})>0\}=\{H_{sm}:sm=1,...,m^+\},
\end{displaymath}
where $m^+\leq m$ and $H_{sm}$ is a relabeling of P-positive sets, and
\begin{displaymath}
\mathcal{H}_m^0=\{H_{rm}:P(H_{rm})=0\}.
\end{displaymath}
A sequence of partitions $\mathcal{H}_m,\mathcal{H}_{m+1},\mathcal{H}_{m+2},...$ is used which is briefly denoted as 
$\{\mathcal{H}_{m+i}:i\geq1\}$ where the notation is adopted
\begin{displaymath}
\mathcal{H}_{m+i}=\{H_{r(m+i)}:r(m+i)=1,2,...,m+i\}
\end{displaymath}
and (see the construction of partitions in subsection 5.4) $\mathcal{H}_{m+i+1}$ is obtained partitioning only one interval
$H_{\underline{r(m+i)}} \in \mathcal{H}_{m+i}$ into two subintervals denoted as
\begin{displaymath}
H_{\underline{r(m+i+1)}},H_{\underline{r(m+i+1)}+1} \in \mathcal{H}_{m+i+1}
\end{displaymath}
with
\begin{displaymath}
H_{\underline{r(m+i+1)}} \cup H_{\underline{r(m+i+1)}+1}=H_{\underline{r(m+i)}}
\end{displaymath}
and including into $\mathcal{H}_{m+i+1}$ all the remaining intervals $\mathcal{H}_{r(m+i)} \in \mathcal{H}_{m+i}$ with
$r(m+i) \neq \underline{r(m+i)}$.Our goal of finding a permutation may be performed assigning to each fixed $n\geq1$ a
corresponding value $E(Y_n) \in \{E(Y_j):j\geq1\}$ such that the convergence holds true
\begin{equation} \label{eq:trentatre}
\lim_{n \to \infty}P_n(H_{rm})=\lim_{n \to \infty}\frac{C_n(H_{rm})}{n}=P(H_{rm}) \textrm{ }\forall H_{rm} \in
\cup_m\mathcal{H}_m.
\end{equation}
The idea of considering the difference 
\begin{equation} \label{eq:trentaquattro}
|\frac{C_n(H_{rm})}{n}-P(H_{rm})|
\end{equation}
is an intuitive one and the assigned value $E(Y_n)$ corresponding to n will be found selecting a set $H_{rm_{0}} \in
\mathcal{H}_m$ and choosing a value $E(Y_{j_{0}}) \in H_{rm_{0}}$;thus we put $E(Y_n)=E(Y_{j_{0}})$.Of course a permutation
has to be found such that the convergence (\ref{eq:trentatre}) holds true $\forall H_{rm} \in \cup_m \mathcal{H}_m$,then 
the possibility is needed of selecting sets inside each $\mathcal{H}_m$ for any fixed $m\geq1$.Moreover the differences
(\ref{eq:trentaquattro}) are not meaningful if the sets $H_{rm} \in \mathcal{H}_m$ are taken when $m>n$:in fact the equality
$C_n(H_{rm})=0$ is trivially satisfied for a large class of $H_{rm} \in \mathcal{H}_m$.Thus a good policy suggests that the 
index m of partitions depends on n,i.e. $m(n)$ is increasing with $m<n$.Recalling that a sequence of partitions 
$\{\mathcal{H}_{m+i}:i\geq0\}$ is used,we assume to work with a strictly increasing sequence of naturals 
\begin{equation} \label{eq:trentacinque}
\{n_{m+i}:i\geq0\}
\end{equation}
and with the sequence of "natural intervals"
\begin{equation} \label{eq:trentasei}
\lbrack n_{m+i},n_{m+i+1})=\{n \in N : n_{m+i} \leq n < n_{m+i+1}\} \textrm{ }\forall i\geq0
\end{equation}
in such a way that for each fixed $n \in \lbrack n_{m+i},n_{m+i+1})$ the selection is performed for a set 
$H_{\overline{r(m+i)_0}} \in \mathcal{H}_{m+i}$ and then we put $E(Y_n)=E(Y_{j_{0}})$ where $E(Y_{j_{0}})$ is a chosen value
of $H_{\overline{r(m+i)_0}}$.Let us observe that when for each $n \in \lbrack n_{m+h},n_{m+h+1})$ we select a set 
$H_{\overline{r(m+h)_0}} \in \mathcal{H}_{m+h}$, at the same time, we still select a set 
$H_{\overline{r(m+i)_0}} \in \mathcal{H}_{m+i}$ for any $i\leq h$:in fact each assigned set $H_{r(m+h)} \in \mathcal{H}_{m+h}$
is a subset ,i.e. $H_{r(m+h)} \subseteq H_{r(m+i)}$ for some $H_{r(m+i)} \in \mathcal{H}_{m+i}$ for any fixed $i\leq h$.
\newline
\underline{2)THE P-NULL SETS}\newline
Given $\mathcal{H}_m$ and its subclass $\mathcal{H}_m^0$ of P-null sets, the union is taken
\begin{equation} \label{eq:trentasette}
B_m^0= \cup \{H_{rm} \in \mathcal{H}_m^0\}.
\end{equation}
Where $n \in \lbrack n_{m+1},n_{m+2})$ and $\mathcal{H}_{m+1}$ is taken,let us describe the set 
$B_{m+1}^0= \cup \{H_{r(m+1)} \in \mathcal{H}_{m+1}^0\}$.The class $\mathcal{H}_{m+1}$ contains the partition into two
subsets $H_{\underline{r(m+1)}}, H_{\underline{r(m+1)}+1}$ of only one set $H_{\underline{rm}} \in \mathcal{H}_m$ and all the
remaining sets $H_{rm} \in \mathcal{H}_m$ with $rm \neq \underline{rm}$. It is now useful to distinguish some cases:\newline
i)if $P(H_{\underline{rm}})=0$, i.e. $H_{\underline{rm}} \in \mathcal{H}_m^0$,then $P(H_{\underline{r(m+1)}})=
P(H_{\underline{r(m+1)}+1})=0$ (because subset of the P-null set $H_{\underline{rm}}$) and $B_{m+1}^0=B_m^0$;\newline
ii)if $P(H_{\underline{rm}}),P(H_{\underline{r(m+1)}}),P(H_{\underline{r(m+1)}+1})$ are all positive,then we have too
$B_{m+1}^0=B_m^0$  because $\mathcal{H}_{m+1}^0$ and $\mathcal{H}_m^0$ contain the same sets.\newline
iii)if $P(H_{\underline{rm}})>0$ and $P(H_{\underline{r(m+1)}})>0,P(H_{\underline{r(m+1)}+1})=0$ (or vice versa
$P(H_{\underline{r(m+1)}})=0,P(H_{\underline{r(m+1)}+1})>0$ ): the class $\mathcal{H}_{m+1}^0$ contains all sets of
$\mathcal{H}_m^0$ and the new set $H_{\underline{r(m+1)}+1}$.Thus  $B_{m+1}^0 \supset B_m^0$ and ,in the general case,we may
write $B_{m+1}^0 \supseteq B_m^0$.\newline
Of course,under iteration of above arguments,we have that $\{B_{m+i}^0:i\geq0\}$ is a non decreasing sequence of P-null
sets where the strict inclusion $B_{m+i+1}^0 \supset B_{m+i}^0$ holds true if the set $H_{\underline{r(m+i)}} \in
\mathcal{H}_{m+i}$,which is partitioned into two subsets $H_{\underline{r(m+i+1)}},H_{\underline{r(m+i+1)}+1} \in
\mathcal{H}_{m+i+1}$,satisfies the same conditions of iii) above,i.e. $P(H_{\underline{r(m+i)}})>0$ and
$P(H_{\underline{r(m+i+1)}})>0,P(H_{\underline{r(m+i+1)}+1})=0$.\newline
\underline{3)THE SELECTION TECHNIQUE}\newline
The technique we consider deals with selection of "next term" $E(Y_{n+1})$ of the permutation,when the first n values
$E(Y_1),E(Y_2),...,E(Y_n)$ are assigned and n satisfies $n_{m+h}\leq n \leq n_{m+h+1}-2$, where h is a fixed natural.Our 
purpose is that of selecting a set $H_{\overline{r(m+h)}_{0}} \in \mathcal{H}_{m+h}$ and then to choose the $(n+1)$-th
value of permutation taking $E(Y_{n+1})=E(Y_j)$,where $E(Y_j) \in H_{\overline{r(m+h)}_0}$.The selection technique is based 
on two different procedures for P-null and P-positive sets.We assume here that any P-null set $H_{rm}$ contains infinitely
many values $E(Y_j)$'s; in fact the case of a P-null set $H_{rm}'$ with finitely many values $E(Y_j)$'s is a trivial one:
the convergence $\lim_{n \to \infty}\frac{C_n(H_{rm}')}{n}=0$ holds true under any permutation.\newline
We consider all partitions $\mathcal{H}_{m+i}$ with $0 \leq i \leq h$,starting with $\mathcal{H}_m$ ,its subclasses
$\mathcal{H}_m^+,\mathcal{H}_m^0$ of P-positive and P-null sets respectively and $B_m^0$ the union of all sets in
$\mathcal{H}_m^0$. A subset is selected 
\begin{equation} \label{eq:trentotto}
N_{m+h} \subset \lbrack n_{m+h},n_{m+h+1})
\end{equation}
and for each $n \in N_{m+h}$ we put $E(Y_n)=E(Y_j)$ where $E(Y_j)$ is a value belonging to $B_m^0$.The choice of $N_{m+h}$
satisfying some conditions which will be discussed later,gives the index values inside $\lbrack n_{m+h},n_{m+h+1})$ where to
place the elements $E(Y_j) \in B_m^0$.Thus if $(n+1) \in N_{m+h}$ we put $E(Y_{n+1})=E(Y_j) \in B_m^0$,while if 
$(n+1) \notin N_{m+h}$ we select a subset $H_{\overline{sm}_0} \in \mathcal{H}_m^+$ using the below method.For each assigned
index $sm_0=1,2,...,m^+$ let us write
\begin{equation} \label{eq:trentanove}
a_{sm_{0}}=|\frac{C_n(H_{sm_{0}})+1}{n+1}-P(H_{sm_{0}})|+\sum_{sm=1,sm \neq sm_0}^{m^+}|\frac{C_n(H_{sm})}{n+1}-P(H_{sm})|
\end{equation}
and define as $\overline{sm_0}$ the index satisfying 
\begin{equation} \label{eq:quaranta}
a_{\overline{sm_0}}=min\{a_{sm_{0}}:sm_0=1,...,m^+\}.
\end{equation}
Recalling that our goal consists in choosing a set inside $\mathcal{H}_{m+h}$,if the selected set $H_{\overline{sm_0}}$
$\in \mathcal{H}_m^+$ is too included into $\mathcal{H}_{m+h}$  we may put $E(Y_{n+1})=E(Y_j)$
where $E(Y_j)$ is a not previously chosen value of $H_{\overline{sm_0}}$.But if $H_{\overline{sm_0}} \notin \mathcal{H}_{m+h}$
,this implies that inside $\mathcal{H}_{m+h}$ there exists a family of sets defining a partition of $H_{\overline{sm_0}}$.
A first partition of $H_{\overline{sm_0}}$ into two subsets may be found inside a class $\mathcal{H}_{m+i1}$ including two
sets denoted by $H_{\underline{r(m+i1)}}$ and $H_{\underline{r(m+i1)+1}}$ such that 
$H_{\underline{r(m+i1)}} \cup H_{\underline{r(m+i1)}+1}=H_{\overline{sm_0}}$ and afterwards a partition of 
$H_{\underline{r(m+i1)}}$ into two subsets may exists inside a class $\mathcal{H}_{m+i2}$ (where $i1<i2 \leq h$) including
two sets $H_{\underline{r(m+i2)}},H_{\underline{r(m+i2)}+1}$ in such a way that $H_{\underline{r(m+i1)}}=
H_{\underline{r(m+i2)}} \cup H_{\underline{r(m+i2)}+1}$.For sake of simplification, and without loss of generality,we may 
suppose that $\mathcal{H}_{m+h}$ contains no further subsets of $H_{\overline{sm_0}}$ than the three subsets
$H_{\underline{r(m+i2)}},H_{\underline{r(m+i2)}+1},H_{\underline{r(m+i1)}+1}$ and the selection of one of the three above
subsets is performed below when all the three subsets have positive P-measure.\newline
The first partition of $H_{\overline{sm_0}}$ into two subsets is introduced by $\mathcal{H}_{m+i1}$;then,after selection 
of $H_{\overline{sm_0}}$,one of the two subsets $H_{\underline{r(m+i1)}}$ or $H_{\underline{r(m+i1}+1}$ is chosen using a
method which is the analogous of above (\ref{eq:trentanove}) and (\ref{eq:quaranta}) when there are only two alternatives.
Thus,given the two quantities
\begin{equation} \label{eq:quarantuno}
b_1=|\frac{C_n(H_{\underline{r(m+i1)}})+1}{n+1}-P(H_{\underline{r(m+i1)}})|+
|\frac{C_n(H_{\underline{r(m+i1)}+1})}{n+1}-P(H_{\underline{r(m+i1)}+1})|
\end{equation}
and
\begin{equation} \label{eq:quarantadue}
b_2=|\frac{C_n(H_{\underline{r(m+i1)}})}{n+1}-P(H_{\underline{r(m+i1)}})|+
|\frac{C_n(H_{\underline{r(m+i1)}+1})+1}{n+1} - P(H_{\underline{r(m+i1)}+1})|
\end{equation}
let us denote by $\overline{k}_0$ the index satisfying
\begin{equation} \label{eq:quarantatre}
b_{\overline{k}_0}=min\{b_1,b_2\}.
\end{equation}
If $b_{\overline{k}_0}=b_2$ then $H_{\underline{r(m+i1)}+1}$ is selected and we put $E(Y_{n+1})=E(Y_j)$,where $E(Y_j)$ is a
not previously chosen value of $H_{\underline{r(m+i1)}+1}$; and this because of the inclusion
$H_{\underline{r(m+i1)}+1} \in \mathcal{H}_{m+h}$.Vice versa,if $b_{\overline{k}_0}=b_1$ the selected set is 
$H_{\underline{r(m+i1)}}$ which is not included into $\mathcal{H}_{m+h}$: in fact $\mathcal{H}_{m+h}$ contains the
two subsets $H_{\underline{r(m+i2)}}$ and $H_{\underline{r(m+i2)}+1}$ of $H_{\underline{r(m+i1)}}$.Then, applying again
(\ref{eq:quarantuno}),(\ref{eq:quarantadue}) and (\ref{eq:quarantatre}) to $H_{\underline{r(m+i2)}}$ and
$H_{\underline{r(m+i2)}+1}$, one of the two sets will be selected; thus we put $E(Y_{n+1})=E(Y_j) \in H_{\underline{r(m+i2)}}$
if $H_{\underline{r(m+i2)}}$ is selected or $E(Y_{n+1})=E(Y_j) \in H_{\underline{r(m+i2)}+1}$ if
$H_{\underline{r(m+i2)}+1}$ is selected.\newline
\underline{4)SELECTING P-NULL SETS} \newline
Recalling the structure of partitions (in the first part of this proof) and considering,for each natural $i\geq0$ fixed,
the interval of naturals $\lbrack n_{m+i},n_{m+i+1})$, our strategy consists in choosing a suitable subset
\begin{displaymath}
N_{m+i} \subset \lbrack n_{m+i},n_{m+i+1})
\end{displaymath}
such that $\forall n \in N_{m+i}$ a value $E(Y_n)$ is selected in such a way that $E(Y_n)=E(Y_j)$ where $E(Y_j)$ is a not
previously chosen value belonging to the set $B_{m+i}^0$ which is the union of all P-null sets inside the partition 
$\mathcal{H}_{m+i}$.As a choice criterion for the set $N_{m+i}$ the following elements are introduced.\newline
Let us consider the family of quotients $\frac{C_n(B_{m+i_n}^0)}{n}$ for each $n\geq n_m$ where $m+i_n=m+i$ 
$\forall n \in \lbrack n_{m+i},n_{m+i+1})$ and $C_n(B_{m+i})$ gives the total number of values in the set 
$\{E(Y_j):j=1,...,n\} \cap B_{m+i}^0$.\newline
The selection of the subset $N_{m+i} \subset \lbrack n_{m+i},n_{m+i+1})$ $\forall i\geq 0$ is performed in such a way that:
\begin{equation} \label{eq:quarantaquattro}
\lim_{n \to \infty} \frac{C_n(B_{m+i_n}^0)}{n}=0
\end{equation}
and
\begin{equation} \label{eq:quarantacinque}
\textrm{all values } E(Y_j) \in \cup_{i=0}^{\infty}B_{m+i}^0 \textrm{are selected.}
\end{equation}
The limit (\ref{eq:quarantaquattro}) above implies the convergence $\lim_{n \to \infty} \frac{C_n(H)}{n}=0$ for each
P-null set $H \in \cup_{i=0}^{\infty} \mathcal{H}_{m+i}$;in fact ,if $H_{r(m+h)}^0 \in \mathcal{H}_{m+h}$ and 
$P(H_{r(m+h)}^0)=0$,we have $H_{r(m+h)}^0 \subset B_{m+h}^0 \subset B_{m+i}^0$ $\forall i \geq h$ and then \newline
$\frac{C_n(H_{r(m+h)}^0)}{n} \leq \frac{C_n(B_{m+i}^0)}{n}$ $\forall i \geq h$; and thus, by limit (\ref{eq:quarantaquattro}),
$\lim_{n \to \infty}\frac{C_n(H_{r(m+h)}^0)}{n}=0$.\newline
We are now ready to choose the next term $E(Y_{n+1})$:if $(n+1) \in N_{m+i}$ we put $E(Y_{n+1})=E(Y_j)$ where $E(Y_j)$ is a
not previously chosen value belonging to $B_{m+i}^0$, while if $(n+1) \notin N_ {m+i}$ we select a P-positive set following
the above procedure and the proof is now complete. $\diamondsuit$\\
As an example/application the extension of Theorem (\ref{eq:cento7}) is suggested to the case of an arbitrary real bounded
and dense sequence $\{t_j:j\geq1\} \subset \lbrack0,T\rbrack$ where each $t_j$ is not necessarily the expectation $E(Y_j)$
of an assigned random variable.Thus the basic elements concerning Theorem (\ref{eq:cento7}) are shown:\\
i) each $t_j$ denotes an observation time of the process $\{Y(t),\forall t \in \lbrack0,T\rbrack\}$ under the assumption
that $t_j \neq t_k, \forall j \neq k $;\\
ii) Q is the counting measure defined over the Borel $\sigma$-field $B\lbrack0,T\rbrack$ such that $Q(A)$ is the total
number of $t_j$'s belonging to A, for each fixed $A \in B\lbrack0,T\rbrack$. Then $Q((a,b\rbrack)= + \infty$ for each
$(a,b\rbrack \subset \lbrack0,T\rbrack$ and $Q(A)$ is a natural value if A is a finite union of points $t \in \lbrack0,T
\rbrack$.\\
iii)The class $\mathcal{M}$ is defined in close connection with Q: position and density of $t_j$'s inside $\lbrack0,T\rbrack$
are elements having a strong impact on $\mathcal{M}$: for instance each absolutely continuous P.M. over $\lbrack0,T\rbrack$
belongs to $\mathcal{M}$. In fact, if P is a P.M. over $\lbrack0,T\rbrack$ with density function $f_P(t)$,each interval H
belonging to any P.S.P. $\{\mathcal{H}_m:m\geq1\}$ of $(0,T\rbrack$ is a P-continuity set and if $P(H)>0 \Rightarrow 
Q(H)=+\infty$ because of the density of $t_j$'s. Thus Theorem (\ref{eq:cento7}) may be applied to any absolutely continuous
measure P over $\lbrack0,T\rbrack$.
\begin{Corollary} \label{eq:one}
If $\{t_j:j\geq1\}$ is a dense subset of $\lbrack0,T\rbrack$,then for each assigned absolutely continuous probability measure P over 
$\lbrack0,T\rbrack$ some permutation $\pi$ can be computed such that the sequence of P.E.M.'s $P_{\pi n}$,which assigns
weight $\frac{1}{n}$ to each point $\{t_{\pi(j)}:j=1,...,n\}$,is vaguely convergent to P.
\end{Corollary}

\end{document}